# Co-optimization of Diffusive and Tomographic Blur in Computed Axial Lithography via Experimental Kernel Identification

Jennings Z. Ye, Abrar Amin Khan, Hayden K. Taylor

Department of Mechanical Engineering, University of California, Berkeley

## Abstract

Computed Axial Lithography is a volumetric additive manufacturing method that selectively cures photosensitive resin through the 3D superposition of patterns of light, offering advantages over layer-based processes including rapid print times, reduced layer artifacts, and compatibility with high-viscosity materials. However, diffusive effects, primarily those of free-radical quenchers such as oxygen, blur the boundary between cured and uncured regions, limiting resolution and preventing the reproduction of sharp, high-spatial-frequency features. By comparing micro-CT data to computational dose models convolved with kernels across a range of diffusivities, we establish a framework for extracting a single diffusion kernel from any standard uncorrected print to account for all observed deviations from the target. In this work, we correct diffusion-induced blurring by co-optimizing for its effects alongside the inherent blur of the computed tomography reconstruction, demonstrating improved fidelity over previous approaches of pre-compensating the target geometry via deconvolution.

## Introduction

Computed Axial Lithography (CAL) is a volume-at-once manufacturing method that selectively cures photosensitive resin through the superposition of light energy. By leveraging tomographic principles, commonly used in medical imaging fields such as computed tomography (CT), CAL patterns light to exceed curing thresholds only in regions forming the target object.[1] For this reason, CAL is also known as Tomographic Volumetric Additive Manufacturing (TVAM).[2] This volumetric approach provides numerous advantages over layer-based processes, such as print times ranging from seconds to a few minutes, a reduction in layer-based artifacts, the ability to use higher viscosity materials, and the capacity to “overprint” around occluded objects.[1–4]

Despite these advantages, CAL faces challenges including material and size limitations due to the need for optical penetration depths comparable to the diameter of the printing volume, the need for manual, geometry-dependent print timing and post-processing, and—central to this work, finite resolution (typically at the 100 µm level in cm-scale parts)[1,2] that limits production of sharp, high-spatial-frequency features. While the diffusion-induced blurring is observed in 2D lithography, where optical proximity correction is standard,[5,6] it is exacerbated in CAL by the limited dose contrast delivered to the material. Unlike in layer-based methods, light in CAL must propagate through regions outside of the target geometry, accumulating background dose. However, even when dose contrast is controlled to acceptable levels, diffusive effects, primarily those of free-radical quenchers such as oxygen, lead to a blurring effect that lowers resolution, particularly at sharp corners or fine features where diffusion path lengths approach the feature size.[1,7–9]

Previous work has corrected for diffusion by pre-compensating the target geometry, deconvolving it with a kernel function derived from the summation of several diffusion kernels, one for each rotation of the vial.[7] This pre-correction is then followed by standard optimization methods that handle the inherent CT blur from the tomographic process. However, serial correction of the two phenomena results in overaccentuated edges and highly non-uniform dose distribution.[10]

In this work, we address this limitation by co-optimizing for diffusive effects alongside the CT blur, leveraging recent advances in iterative optimization methods for CAL.[10,11] Here, we experimentally derive a single 3D kernel for the entire printing and post-processing procedure by comparing uncorrected test prints to computational models of the delivered dose convolved with different blurring kernels. While we fit a notional diffusion coefficient using our initial experimental results, we additionally test correction levels across a range of diffusivities. This allows us to evaluate the effectiveness of both the correction model and the diffusivity determination method, as well as the sensitivity of the correction method to this diffusion coefficient.

# Results

## Correction method

In this work, we use a single 3D blurring kernel to model the cumulative effects causing discrepancies between the delivered dose and the printed object. While our model accepts any kernel function, we opt for a diffusion kernel due to the well-documented role of diffusion in acrylate-based printing, as well as the simplicity of fitting a kernel that requires only the diffusion coefficient alongside a known print duration[1,7–9]. Specifically, we use the following kernel (Eq. 1), which describes the steady-state concentration distribution that develops around a small cubic sink of a species with diffusivity $D$.[12]

$$K(x,y,z) = \int_{-\frac{d}{2}}^{\frac{d}{2}}\int_{-\frac{d}{2}}^{\frac{d}{2}}\int_{-\frac{d}{2}}^{\frac{d}{2}} \frac{1}{4\pi D\sqrt{(x-x')^2+(y-y')^2+(z-z')^2}} \operatorname{erfc}\left(\frac{\sqrt{(x-x')^2+(y-y')^2+(z-z')^2}}{2\sqrt{Dt}}\right) \mathrm{d}x'\,\mathrm{d}y'\,\mathrm{d}z' \quad (1)$$

where $D$ is the diffusion constant

$t$ is the total print duration

$d$ is the voxel size for the discretization of the object

In the standard filtered-back projection (FBP) algorithm[13], the projection is defined by the Radon transform (Eq. 2)[14] followed by filtering with a Ram-Lak (ramp) filter with potential additional window functions[15,16]. To be physically realizable, the projections must be transformed to be non-negative. Eq. 3 shows the full formulation for generating this initial projection.

$$Rf(r,\theta,z) = \int_{L(r,\theta,z)} f(x,y,z)ds \quad (2)$$

$$g_0 = \max\left(0, \mathcal{F}^{-1}\{H(\omega)\cdot\mathcal{F}\{Rf(r,\theta,z)\}\} + b\right) \quad (3)$$

where $f$ is the target dose distribution

$L(r,\theta,z) = \{(x,y,z) \in \mathbb{R}^3 \colon x\,cos\theta + y\,sin\theta = r\}$

$g$ is a projection set, with initial projections $g_0$

*b* is an intensity offset, set to 0 in this work to enforce non-negativity via direct clamping, consistent with Projected Gradient Descent (PGD)

$H$ is the frequency-domain Ram-Lak (ramp) filter

$\mathcal{F}$ denotes the 1D Fourier transform along dimension $r$

Inversion of the Radon transform to generate reconstructions is performed by applying the adjoint operator (back-projection) (Eq. 4) to the filtered projections. Due to the information loss inherent in the non-negativity constraint, iterative optimization of the projections is used to improve the resolution and contrast of the reconstruction.

$$R^*g(x,y,z) = \int_0^{\pi} g(r,\theta,z)d\theta \quad (4)$$

To account for diffusion, we model the physical dose deposition (Eq. 5) by convolving the standard FBP reconstruction with the kernel from Eq. 1. The corresponding adjoint operator (Eq. 6), required for the gradient calculation, is derived by applying the adjoint kernel prior to the Radon transform.

$$\tilde{R}^*g(x,y,z) = R^*g(x,y,z) * K(x,y,z) \quad (5)$$

$$\tilde{R}f(r,\theta,z) = \int_{L(r,\theta,z)} f(x,y,z) * K^*(x,y,z)ds \quad (6)$$

Note that $K^* = K$ for a spherically symmetric kernel function such as our diffusion kernel.

$*$ represents a convolution, computationally implemented using multiplication in Fourier space.

We then perform iterative optimization of the projections using the Band-Constraint-$L_p$ norm (BCLP) optimization scheme,[11] under which the gradient for co-optimization of diffusion alongside the standard CT blur can be calculated by substituting the standard forward and backward operators with their modified versions defined in Eqs. 5 and 6.

## Determination of Kernel Parameters

To extract a single 3D kernel that accounts for deviations between the printed object and the target, we first print a set of uncorrected objects by voxelizing a target geometry from an STL file and generating projections and reconstructions using the standard Radon transform, FBP, and projected gradient descent (PGD). Assuming the diffusion coefficient is material-dependent and geometry-independent, we use two geometries throughout this work. Since CAL struggles to produce sharp corners, we select a plate structure with 500 µm positive and negative square features at different radial positions, as well as a round through-hole for comparison. The second geometry is a three-column structure involving cylindrical 600 µm diameter pillars capped with flat rectangular plates, chosen to highlight deviations between pillars at various radial positions and transitions between the pillars and plates. Figure 1 shows the STL geometry files along with optical micrographs of three uncorrected samples. All objects in this work were printed using pentaerythritol tetraacrylate with 10 mM camphorquinone, an equal mass of ethyl 4-dimethylaminobenzoate, and 1 mM TEMPO.

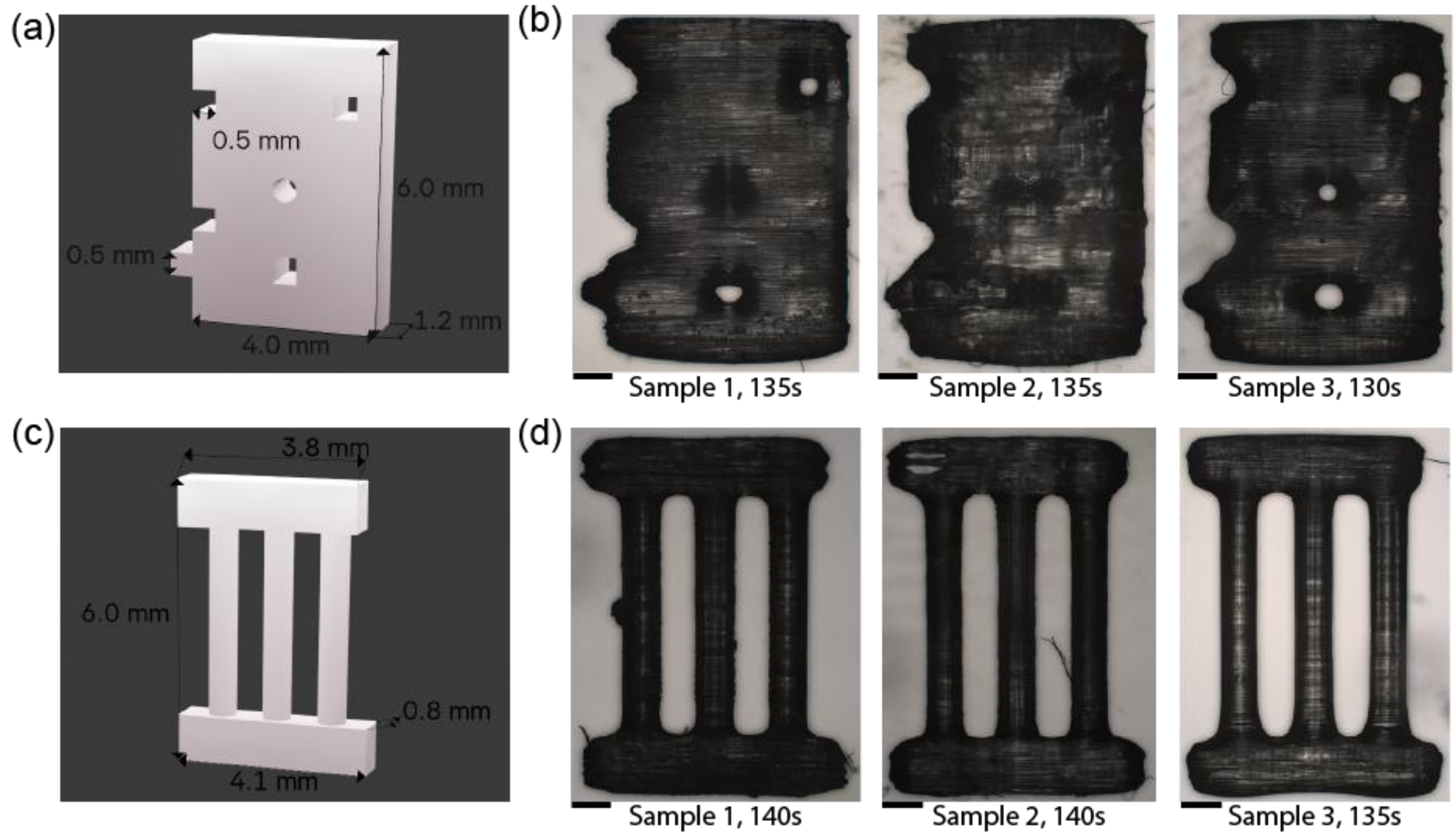

**Figure 1.** (a) Plate geometry STL mesh, units in mm. (b) Optical micrographs of uncorrected prints for the plate geometry. 600 µm scalebars. (c) Three-column geometry STL mesh, units in mm. (d) Optical micrographs of uncorrected prints for the plate geometry. 600 µm scalebars.

Visual inspection reveals the blunting of sharp features, with all positive and negative square features appearing rounded. In addition to visual inspection, we performed micro-CT scans for quantitative comparison of the printed objects to the target STL model. To extract the material diffusivity, we computationally reconstructed the geometry across a range of diffusion coefficients by convolving the diffusion-free dose with different kernels. The diffusion-incorporated doses were then binarized at thresholds that yielded a cured volume equal to that of the target. Figure 2 illustrates this binarization process.

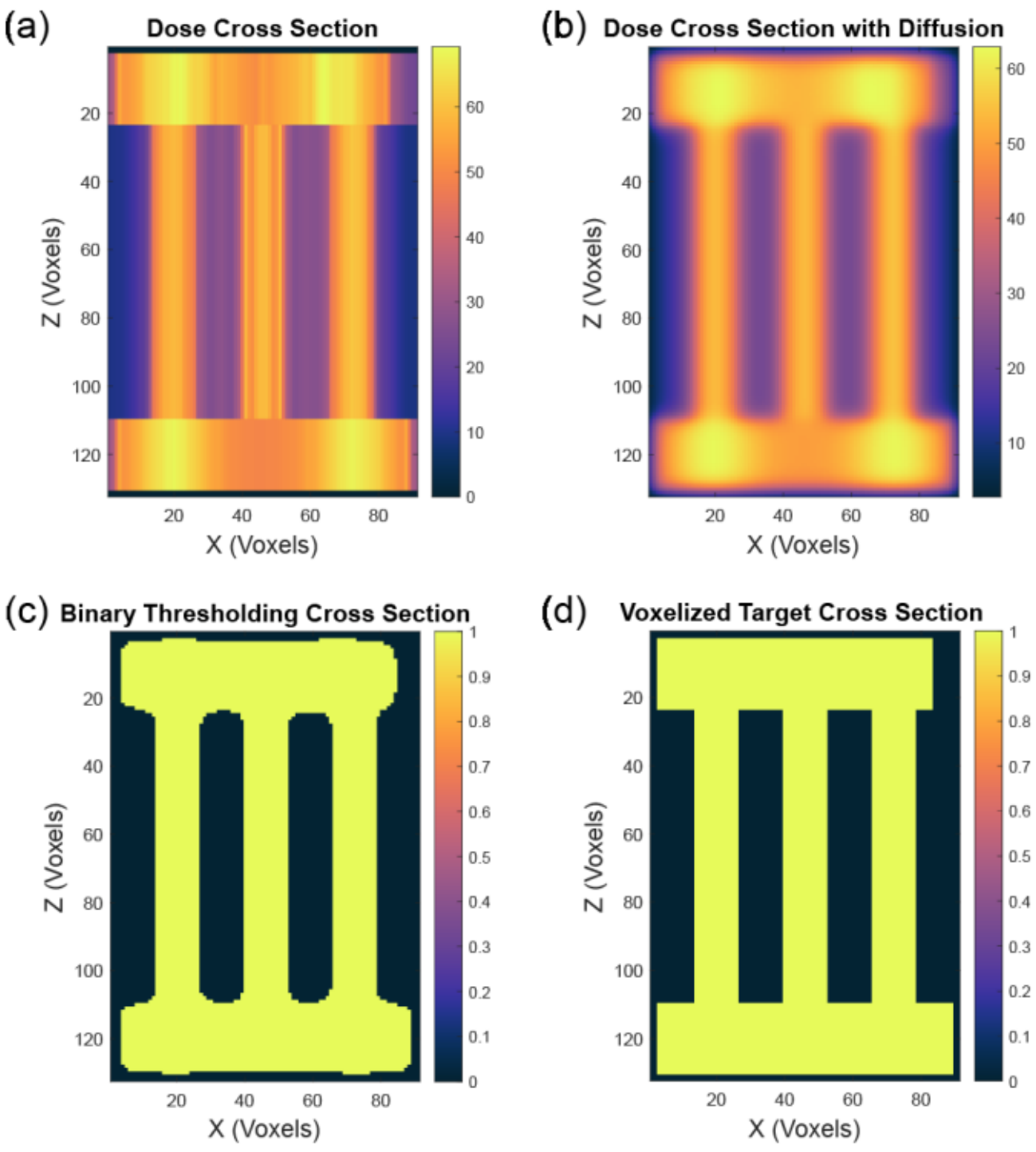

**Figure 2.** (a) Dose reconstruction slice from standard FBP and projected gradient descent. (b) Dose reconstruction slice from the same projections with diffusive effects considered, kernel inputs $D = 1 \times 10^{-10}\ m^2/s, t = 140\ s$. (c) Thresholded binary reconstruction slice with diffusion. (d) Voxelized target slice. 47 µm voxel size throughout.

Micro-CT data were registered to the voxelized target and compared to computational reconstructions across a range of diffusion coefficients using two metrics. The first, average symmetric surface distance (ASSD), is the average distance between surface voxels of the scan data and the voxelized target. This provides a global boundary error in physical units, where a smaller distance represents lower error. Since our structures, particularly the plate, are largely planar, small systemic out-of-plane deviations can dominate boundary-based error metrics. To address this, we additionally compute the multiscale structural similarity index measure (MS-SSIM), where a higher score indicates greater similarity, against 2D thickness maps that represent the number of cured voxels perpendicular to the plane of the structure. Figure 3 displays these metrics for different modeled diffusivities, as well as examples of the 2D thickness maps.

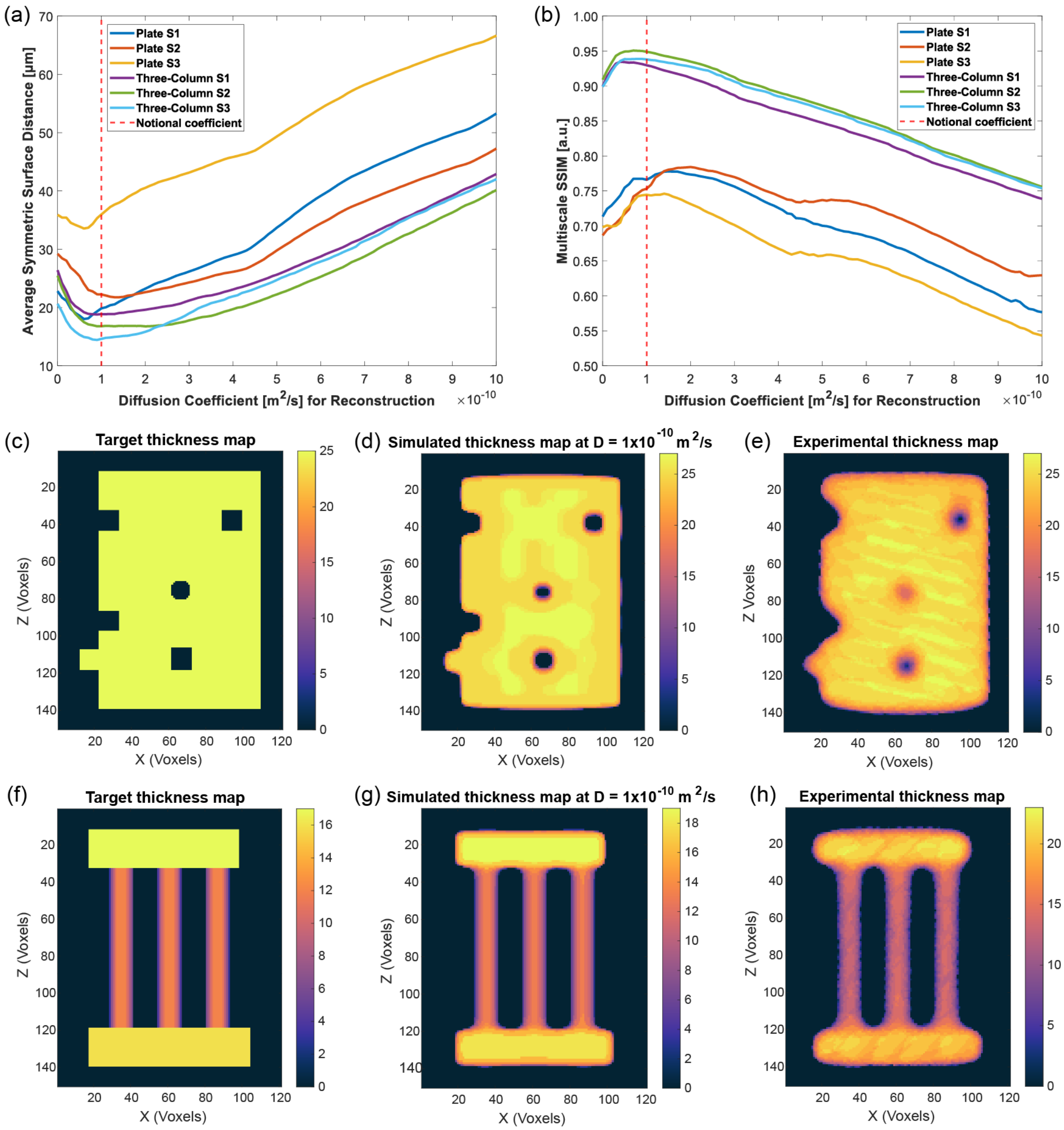

**Figure 3.** (a) Average symmetric surface distance (ASSD) across different simulated diffusion coefficients. (b) Multiscale structural similarity index measure (MS-SSIM) across different simulated diffusion kernels. (c) Thickness map of voxelized plate target for MS-SSIM and visualization. (d) Thickness map of binarized simulation of uncorrected plate projections at a notional diffusion coefficient $D = 1 \times 10^{-10}\ m^2/s$. (e) Experimental thickness map for sample 1 of the uncorrected plate geometry, from voxelized micro-CT data after registration to the voxelized target. (f–h) Thickness maps of voxelized target, simulated binarization, and experimental result (sample 1) of the three-column geometry. 47 μm voxel size throughout.

From the plots in Figure 3, we determined a notional diffusion coefficient as the value that yields close to the lowest ASSD or highest MS-SSIM across our six samples. Note that our simulated binarized reconstructions used for fidelity metrics are of a constant volume equivalent to that of the target (Fig. 2). However, even with experimental timing control, the degree of over- or under-curing varies between samples. Since the dataset includes both visually over-cured (Plate S2) and under-cured (Plate S3) prints, we can nonetheless analyze the full dataset to determine a notional diffusion coefficient.

If chemical diffusion is the dominant effect, we expect the results to be largely geometry-independent. While the optimal diffusivity varies across samples and metrics — most notably, the plate geometry exhibits its best MS-SSIM at a higher diffusion coefficient than the three-column geometry — all samples are near their optimal value at $D = 1 \times 10^{-10}\ m^2/s$. We therefore adopt this as a notional coefficient intended to capture the effects of all relevant mass diffusion across the remainder of our analysis. The Dice similarity coefficient, Jaccard index, and 95th percentile Hausdorff distances were also computed and exhibited results consistent with ASSD (S.1).

## Corrected projections and prints

To both assess the validity of our diffusivity fitting process and investigate the sensitivity of the correction to changes in the diffusion coefficient, we generated several sets of modified projections for each geometry: one at our notional diffusion coefficient of $D = 1 \times 10^{-10}\ m^2/s$, and three at higher, "overcorrected" values of $D = 2 \times 10^{-10}\ m^2/s$, $D = 3 \times 10^{-10}\ m^2/s$, and $D = 6 \times 10^{-10}\ m^2/s$. Kernel time inputs were derived from the experimental print durations of our initial uncorrected tests, multiplied by the dose delivery ratio between the initial and diffusion-corrected projections (S.2).

We performed three prints at the notional diffusivity of $D = 1 \times 10^{-10}\ m^2/s$ and one at each of the higher values. As with the initial uncorrected samples, micro-CT scans and optical micrographs were used to evaluate the fidelity of these prints. Computational models generated using these projections assume the true diffusion coefficient matches our notional value of $D = 1 \times 10^{-10}\ m^2/s$. Figures 4 and 5 depict computational thickness maps alongside the micro-CT results and optical images for three of the correction levels. Optical images of the remaining samples can be found in S.3.

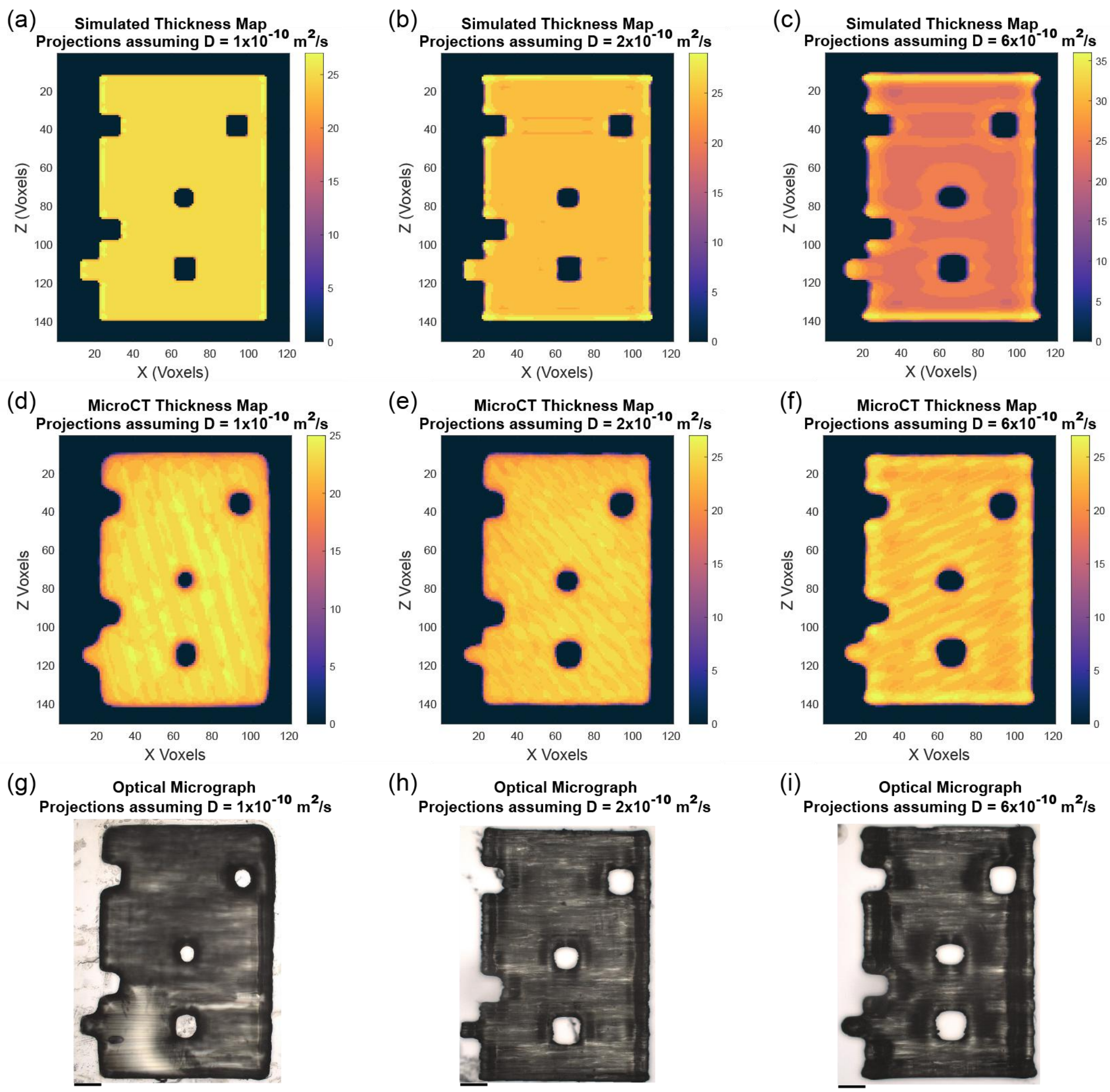

**Figure 4.** Simulated and experimental results for the corrected plate projections. (a–c) Simulated thickness maps for different correction levels, assuming a true coefficient of $D = 1 \times 10^{-10}\ \mathrm{m}^2/s$. (d–f) Thickness maps from voxelized micro-CT scan data of experimental samples using projections optimized for different diffusion coefficients. (g–i) Optical micrographs of the printed samples. Scalebars: 600 µm.

Visual inspection indicates that all correction levels resolve the 500 µm holes of the plate more effectively than the uncorrected samples. However, while sharp corners are improved, they remain notably rounded. Additionally, compared to simulations, the "overcorrected" results do not exaggerate edges to the extent predicted by simulation. In particular, the result at $D = 2 \times 10^{-10}\ m^2/s$ exhibits sharper corners than the result at $D = 1 \times 10^{-10}\ m^2/s$ without displaying signs of overcorrection. Diagonal periodic striations seen in the micro-CT data do not correspond to features in the optical images and are likely systematic noise from the scanning process itself. In contrast, horizontal striations visible in optical images are known artifacts resulting from well-documented “self-writing waveguide” effects during light propagation CAL.[17,18]

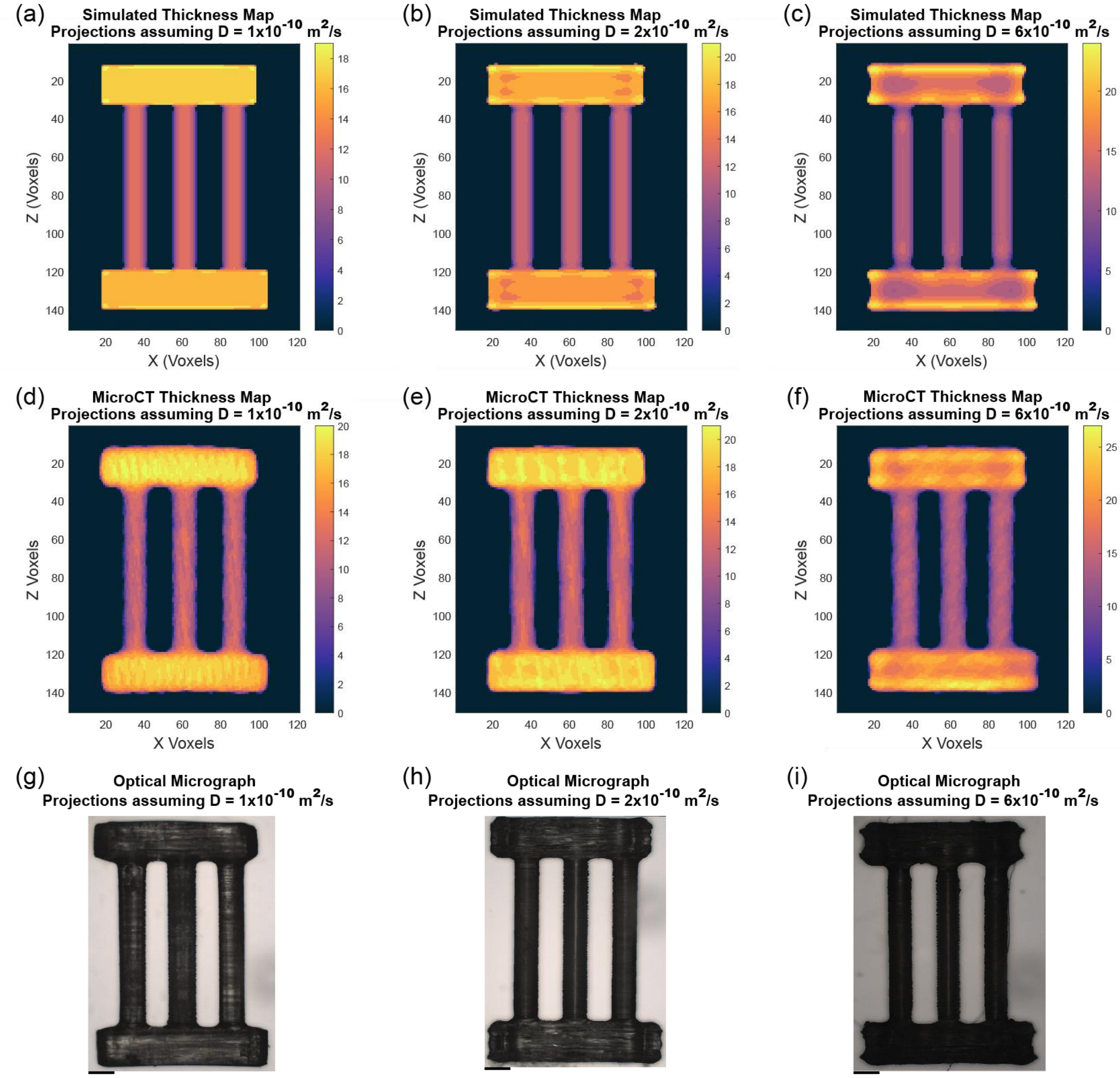

**Figure 5.** Simulated and experimental results for the corrected three-column projections. (a–c) Simulated thickness maps for different correction levels, assuming a real coefficient of $D = 1 \times 10^{-10}\ m^2/s$. (d–f) Thickness maps from voxelized micro-CT scan data of experimental samples using projections optimized for different diffusion coefficients. (g–i) Optical micrographs of the printed samples. Scalebars: 600 µm.

For the three-column geometry, corrected projections again resolve the square edges and sharp transitions better than uncorrected results, but do not achieve the 90° angles of the design. Compared to the plate structure, the projections generated assuming $D = 6 \times 10^{-10}\, m^2/s$ demonstrate overcorrection more clearly, as the edges, particularly in the optical image, are flared outward with erroneous concave profiles. Additionally, the nonuniformity in the top and bottom plates for this projection is evident in the micro-CT thickness maps, further indicating that this projection set overdelivers dose to the edges. Comparisons between the three columns reveal no observable differences, indicating that there was no meaningful radial dependency of the diffusivity in our experiments.

In addition to visual inspection, fidelity at all correction levels was evaluated using the ASSD and MS-SSIM metrics by registering both computational reconstructions and micro-CT data to the original voxelized target. To account for the volumetric shrinkage of acrylates during curing, the registration process also solved for an isotropic scaling factor (discussed in S.4). Figure 6 shows ASSD and MS-SSIM metrics for each correction level, alongside a composite metric that combines the two measures across both geometries for reduced noise. To evaluate the combined metric, we scaled the ASSD and MS-SSIM scores for each geometry to the [0, 1] range, where 0 and 1 represent the worst and best achieved score for each geometry across any corrected diffusivity level. These scaled scores were then averaged to obtain the final composite metric. A quadratic fit was used to examine the trends between achieved fidelity and the correction level. While the true shape of the fidelity curve is difficult to ascertain, the expectation of a single peak when corrected at the true diffusivity with a falloff on either side makes the quadratic form a useful local approximation.

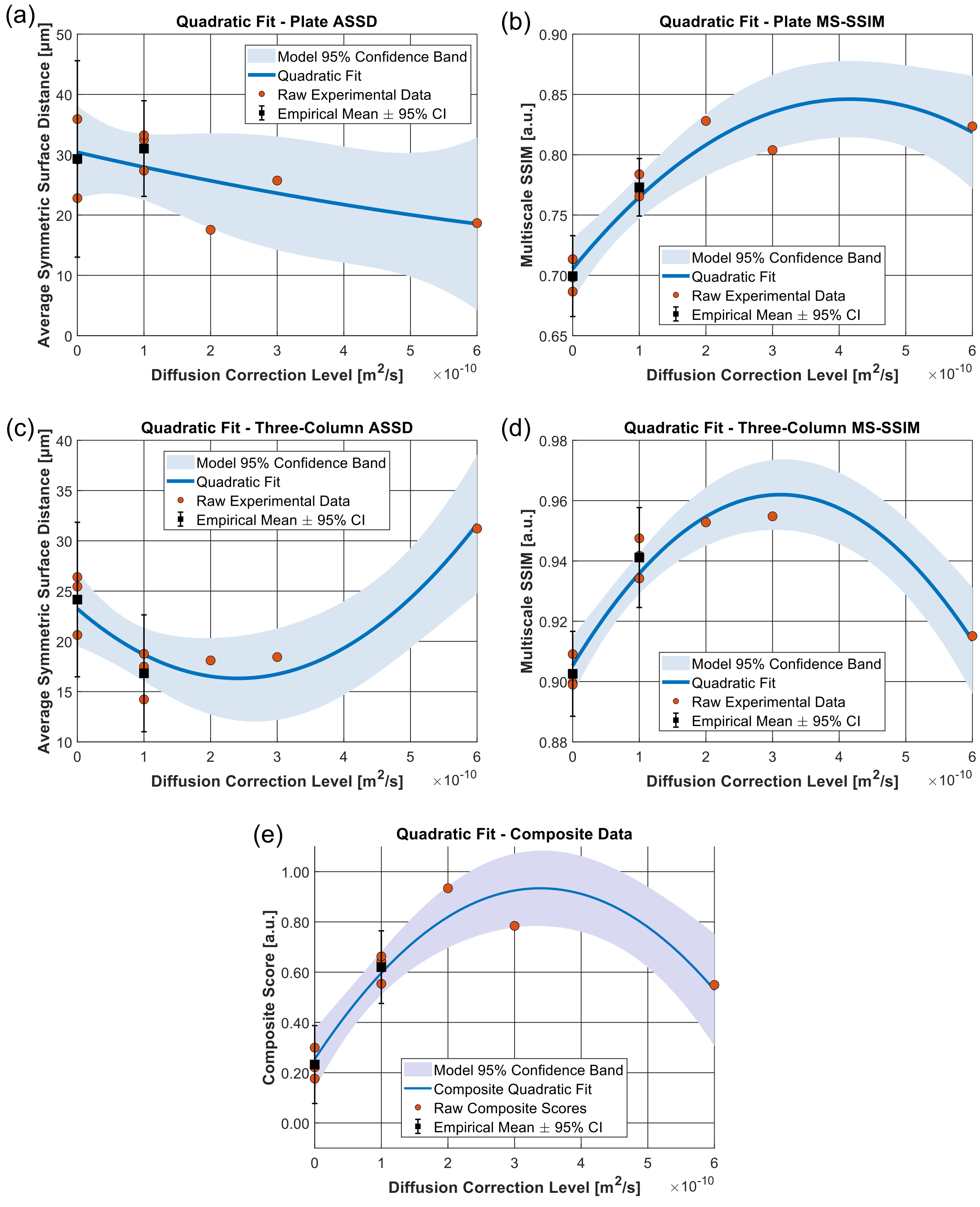

**Figure 6.** Quantitative experimental fidelity metrics for corrected prints. (a) Plate ASSD. (b) Plate MS-SSIM. (c) Three-Column ASSD. (d) Three Column MS-SSIM. (e) Composite score, taken by averaging scaled scores for ASSD and MS-SSIM across both geometries.

Apart from the ASSD metrics for the plate, the quantitative results show improvement over the uncorrected prints at correction levels of $D = 1 \times 10^{-10}\, m^2/s$, $D = 2 \times 10^{-10}\, m^2/s$, and $D = 3 \times 10^{-10}\, m^2/s$ on average. These results appear statistically significant in the MS-SSIM results but are less conclusive in the ASSD data. Consistent with visual inspection, the quantitative metrics suggest that our notional coefficient of $D = 1 \times 10^{-10}\, m^2/s$ may be an underestimate, with the composite score achieving peak fidelity at $D = 3.4 \times 10^{-10}\, m^2/s$, again with the caveat that the quadratic form may have difficulties in fitting the true shape of the curve. As also suggested in the original fitting of the diffusion coefficient, the plate seems to perform comparatively better than the three-column structure at higher values of $D$, although more samples across the diffusivity range would be needed to draw more conclusive data about different trends across the geometries. Overall, a wide range of correction levels yield significant improvement over the uncorrected baseline.

## Comparisons to pre-correction via deconvolution

Our correction method leverages recent advances in CAL projection optimization to incorporate the diffusion convolution directly into the gradient of an analytical loss function, co-optimizing for diffusion alongside the standard CT blur. In contrast, the previous approach demonstrated by Orth *et al.* deconvolved the target dose with a diffusion kernel as a separate pre-compensation step prior to optimization.[7] In that framework, the diffusion-free dose was optimized against the deconvolved target with the aim that the final printed object would achieve the target dose profile after diffusion. Because the loss function in our approach is defined directly on the convolved reconstruction, remaining errors in the optimized result are prevented from further propagation through convolution with the kernel, as occurs in the pre-compensation method.

Here, we conduct a comparison by performing a standard Richardson-Lucy (RL) deconvolution of the voxelized target with the diffusion kernel at the notional $D = 1 \times 10^{-10}\, m^2/s$. Projections were then generated from this pre-compensated target using standard FBP followed by iterative optimization using the same BCLP method and parameters as our co-optimization scheme. A comparison of results for the plate geometry is shown in Figure 7.

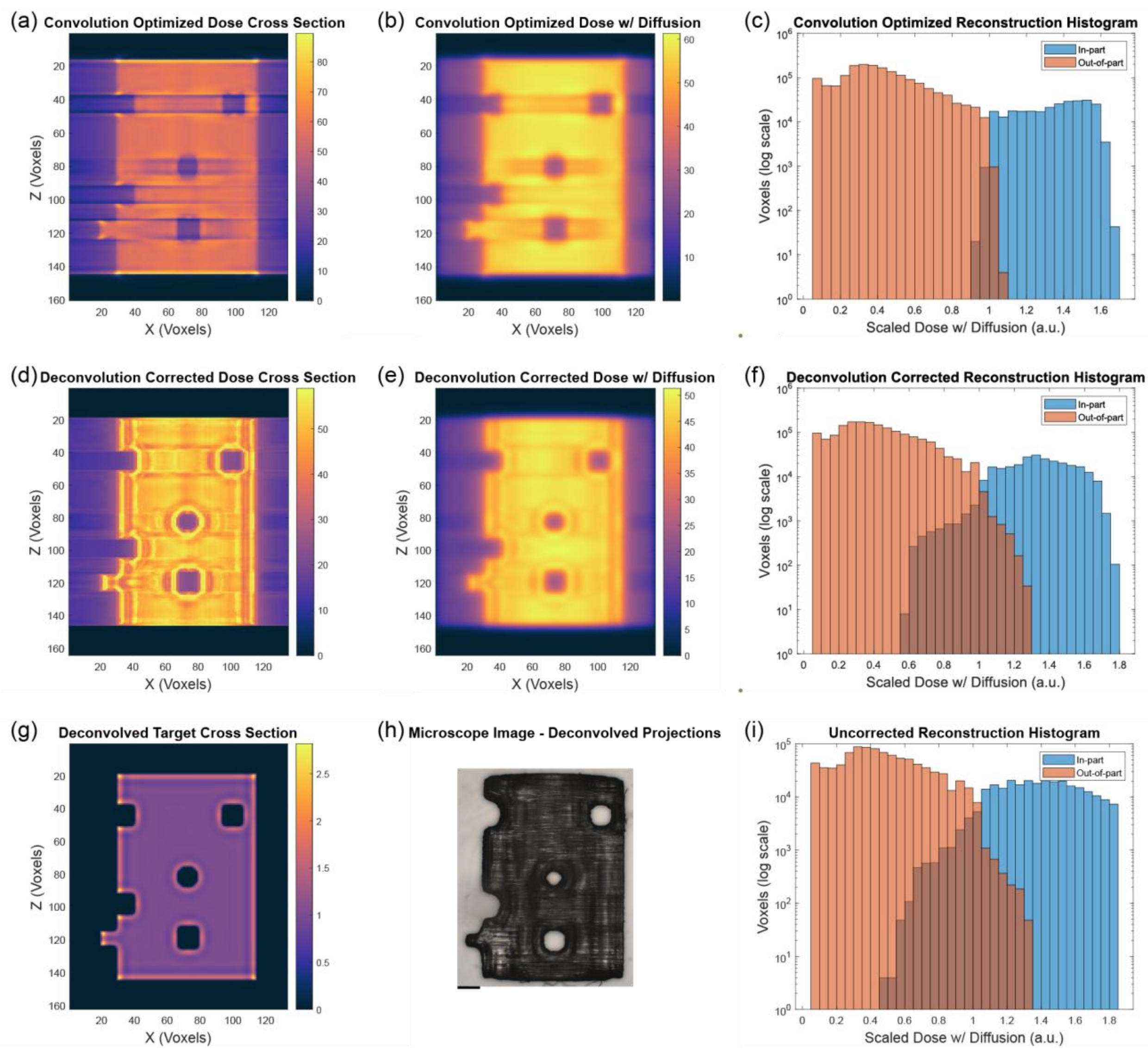

**Figure 7.** (a) Dose cross-section without diffusion from our convolution optimized method. (b) Dose cross-section with diffusion from our method. (c) Histogram of in- and out-of-part voxel doses for our method. (d) Dose cross-section without diffusion from deconvolution method of Orth *et al.* (e) Dose cross-section with diffusion from deconvolution method. (f) Histogram of in- and out-of-part voxel doses for the deconvolution method. (g) Deconvolved target dose. (h) Optical micrograph of printed object using projections from deconvolution method. Scalebar: 600 µm. (i) Histogram of voxel doses for uncorrected projections for comparison. Histograms are scaled so that curing all voxels with scaled dose greater than 1 results in a volume equal to that of our voxelized target. All results corrected for $D = 1 \times 10^{-10}\ m^2/s$.

Both simulated and experimental results indicate that the direct deconvolution method tends to generate more artifacts by overaccentuating the edges and underexposing adjacent regions, consistent with previous findings.[10] Comparing with the results previously shown in Figure 4, the co-optimization scheme yields greater uniformity across interior regions in both simulations and experiments. Histogram analysis further reveals that the deconvolution method shows a significantly higher overlap between in-part and out-of-part voxel doses compared to our method, more closely resembling the distribution of the uncorrected projections.

Quantitatively, the deconvolution method yielded an experimental ASSD (42.5 µm) that was inferior to, and MS-SSIM (0.696) comparable to, that of the uncorrected samples. Although this projection set resolved all holes with visual conformity comparable to our method, overall quantitative fidelity was limited, likely due to local surface deviations near sharp features.

# Discussion

In this work, we demonstrate a method for sharpening of features in CAL by correcting for well documented diffusive effects[1,7,8] using recently developed optimization methods combined with a single experimentally determined kernel. In implementing our diffusion kernel, we assume an infinite domain as our printing parameters leave more than a millimeter of margin between the target geometry and the vial edges—a distance significantly larger than the diffusion effects observed in our results. In addition, the kernel implementation assumes that the volume rotates sufficiently quickly to model each voxel as receiving a time-averaged continuous dose rather than a transient one. At a 47 µm voxel size and a rotation rate of 36°/s (6 RPM), the diffusion timescale for a single voxel is approximately $\frac{(4.7*10^{-5}\mathrm{m})^2}{10^{-10}\mathrm{m}^2/s} \approx 22\ s$ using our original notional diffusivity of $1\ \times 10^{-10}\ \mathrm{m}^2/\mathrm{s}$, or about 6.5 seconds assuming the optimal diffusivity of $3.4\ \times 10^{-10}\ \mathrm{m}^2/\mathrm{s}$ extracted in Figure 6. These diffusion timescales are modestly greater than the five-second periodicity with which the illumination intensity at a given point in the material cycles during printing (assuming the material is highly transmissive, the intensity repeats every 180° of rotation). In addition, the dose delivered to a position generally varies smoothly during rotation, reducing the transient dose variation to the voxel. For these reasons, the use of time-averaging throughout the printing process seems a reasonable approximation. However, future work could investigate the ability of time-resolved simulations to improve upon these model assumptions.

While we determined a single notional diffusion kernel across two geometries based on the assumption of geometry independence, our process revealed small variations in the diffusion coefficient that best matched our initial tests (Fig. 3). Additionally, corrected results across a range of diffusion coefficients revealed a potential underestimation of the diffusivity. However, our extracted notional radical quencher (e.g., oxygen) diffusivity of $D = 1 \times 10^{-10}\ m^2/s = 1 \times 10^{-4}\ mm^2/s$ is comparable to values ( $D = 1.51 \times 10^{-4}\ mm^2/s$ and $D = 0.6 \times 10^{-4}\ mm^2/s$ ) reported by Orth *et al.*[7] for other acrylates (DUDMA/PEGDA and DUDMA, respectively). While the aim here was to extract a correction kernel using a single set of uncorrected prints, future work can refine the diffusion coefficient through additional prints in a similar process as demonstrated in Figure 6. Alternatively, a second kernel could be fit to results printed at the initial notional diffusivity and analytically combined with the original kernel in an iterative process.

Our process provides a generalized framework fitting any kernel form using uncorrected printed parts from any sample geometry. While we used a diffusion kernel here, future work may explore other kernel types, such as a diffusion-reaction kernel[19] or any other kernel that may capture specific process dynamics. Additionally, although we use a single spherical kernel in object space, this correction could be combined with other corrections in both projection and object space, such as the optical point spread function (PSF), by updating the gradient calculation. We assumed a perfectly sharp optical PSF due to our small pixel size of 47 µm relative to the deviations seen in the uncorrected prints. The projections used in this work do, however, include an absorption correction (S.5), although the effects are minimal for our highly transparent material across the small 6 mm vials.

Since CAL prints are timed manually, fidelity is heavily dependent on print timing and the degree of over- or under-curing. With a modified reconstruction that accounts for diffusion and more closely resembles the object, we expect relative dosing rates to track print time more accurately; however, predicting print times remains difficult without an initial reference (S.2). Further investigation of the relationship between gelation thresholds, dosing rates, and diffusion can be crucial for temporal control in CAL. Higher dosing rates both decrease print time and reduce the effects of diffusion, potentially sharpening results. Since targeted reconstructed dose thresholds are somewhat arbitrary in CAL and are often set based on the initial reconstruction[10,20], future studies may investigate maximizing projection intensity to minimize print time, bounded by the contrast limits imposed by saturation (S.2). Additionally, loss functions may be modified to have a penalty term for lower intensities or to decouple the gradient contributions of the diffusion coefficient from those dependent on print time or dosing rate.

In this work, equal-volume thresholding of the reconstructions was used to determine notional diffusivity, projection dosing rates, and the subsequent kernel input times (S.2). However, volume preservation does not strictly yield minimum geometric error. Registering experimental results to reconstructions at different thresholds could enable better analysis of the model fit by accounting for over- or under-curing of the printed object, though running 3D registration across a range of unknown thresholds is computationally expensive.

Significant progress has been made recently in improving the print fidelity of CAL through both finer optics[21] and improved projection generation and modeling [7,8,16,22]. Our tests demonstrate a method of diffusion correction that yields improvements in both the corner fidelity of larger bulk regions and the ability to resolve small 500 µm features. This capability may be of particular interest when applying CAL to higher-diffusivity materials, such as hydrogels.[23] Beyond edge sharpening, our method may enable the printing of features at length scales not resolvable with standard FBP projection generation. Geometries enabled by this diffusion compensation could include internal flow channel networks and lattices. Ultimately, our research presents another step towards enabling CAL for a wide variety of use cases by providing a robust framework for the modeling and correction of diffusive effects within the printing material.

# Methods

## Optical setup

We conducted prints using an LED-based CAL setup similar that reported by Bhattacharya *et al.*[4] The light source has a specified wavelength of 455 nm (measured to be 459 nm). Samples were printed in 6 mm diameter vials, set to rotate at 36°/s and submerged in refractive index (RI)-matching box of mineral oil $(RI{\sim}1.48)$. For in-situ monitoring of the printing process, a shadowgraph system was used as in Bhattacharya *et al*.[4]

## Material formulation and post processing

Pentaerythritol tetraacrylate (PETA) was used as the monomer due to its well documented properties and high transmission (99.9%) at 455 nm wavelength across a 1 cm path length. PETA was mixed with 10 mM camphorquinone (CQ), an equal mass of ethyl 4-dimethylaminobenzoate (EDAB), and 1 mM of 2,2,6,6-Tetramethylpiperidine 1-oxyl (TEMPO) to create the photoresin. TEMPO was used as a radical scavenger and CQ as the photo-inititator, with EDAB acting as the activator for CQ. The photoresin was stored in a dark environment at room temperature. The material crystallized at room temperature if stored for several hours. Thus, the material was heated to 60 °C until the resin was entirely transparent and subsequently cooled to room temperature before printing to maintain temperature conditions across experiments.

After the printing process, the object was kept inside the vial and rinsed with propylene glycol methyl ether acetate (PGMEA) and then developed in isopropyl alcohol (IPA) for 5–10 rinses until all uncured resin are visibly drained from the vial. The objects were then carefully transferred to glass slides and left to dry at room temperature for characterization.

## Projection generation and optimization

Forward and backward projection functions as well as filtering of the projections were performed using our open-source release of the software: the CAL-software-Matlab repository on Github[1] and ASTRA Toolbox[24,25]. The modified Radon and adjoint Radon functions (Eq. 5 and 6) were implemented by performing FFT-based convolutions where necessary. Eq. 6 with the added filtering and clamping step from Eq. 2 was also used in generating the initial guess for the projections. Kernel generation details as well as alternative methods of generating the initial guess can be found in S.6.

Optimization of the projections using the BCLP loss function and gradient[11], using a standard 2-norm, are shown below. Substituting the results of Eq. 5-6 into the loss function defined by BCLP, we get the following.

$$E(\underline{\boldsymbol{r}}) = \max\left(0, \left|\left((\tilde{R}^* g)(\underline{\boldsymbol{r}})\right) - f_T(\underline{\boldsymbol{r}})\right| - \varepsilon(\underline{\boldsymbol{r}})\right)$$

$$\mathcal{L} = \left(\int_V w(\underline{\boldsymbol{r}})\, E(\underline{\boldsymbol{r}})^2\, d\underline{\boldsymbol{r}}\right)^{\frac{1}{2}} \tag{7}$$

$$\nabla_g \mathcal{L}(\underline{\boldsymbol{r}}') = \mathcal{L}^{-1} \tilde{R}\left(w(\underline{\boldsymbol{r}})\, E\, sgn\left(\left((\tilde{R}^* g)(\underline{\boldsymbol{r}})\right) - f_T(\underline{\boldsymbol{r}})\right) * K^*(\underline{\boldsymbol{r}})\right)(\underline{\boldsymbol{r}}') \tag{8}$$

Where $\underline{\boldsymbol{r}}$ represents spatial points

$\underline{\boldsymbol{r}}'$ represents sinogram points

$E$ is the band constraint error of each voxel with some tolerance band $\varepsilon$

$w$ are user defined weights for each spatial coordinate (voxel), set to a constant in this work

$f_T$ is the target dose

$sgn$ denotes the sign (signum) function

BCLP code was custom-written for MATLAB following the above. Our implementation includes a few modifications, details of which can be found in S.7.

## Micro-CT and Fidelity Quantification

Micro-CT scans were performed using a General Electric eXplore Locus micro-CT at a 46.9 µm detector spacing. 3D Slicer[26] was used to convert DICOM data into STL meshes, applying median denoising in the 2,2,2 neighborhood and auto-thresholding using Otsu's method[27].

These meshes were binarized and registered to the original target STL using MATLAB's "RegularStepGradientDescent" optimizer. Since the objects' designs have large bulk regions, the optimizer can be prone to getting stuck in local minima. As a result, eight initial orientations were given to the optimizer to run in parallel, with the best result taken and the orientation verified via visual inspection. These registered volumes were then used to compute the ASSD (using "bwprim" and "bwdist") and MS-SSIM ("multiSSIM3") against the voxelized target of the original STL, as well as against the simulated results at different diffusion constants. As MS-SSIM is sensitive to padding, all objects for comparison were padded to a size of [121, 121, 150]. Quadratic fits and confidence intervals to the quantitative metrics were obtained using MATLAB's "fitlm" and "predict" functions.

# Data and Code Availability

All data supporting the findings of this study are available within the paper and its supplementary information files. The code used for projection generation and fidelity evaluation will be made publicly available on GitHub at a later date.

# Author Contributions

J.Y. and A.K. contributed equally to this work. J.Y. wrote the code and generated the projections, helped with experiments and quantitative analysis, and led in writing. A.K led printing, post-processing, and characterization via microscopy and micro-CT and contributed to writing and reviewing. H.T. conceptualized, supervised, and reviewed the work.

# Supplementary Information

## S.1 Other metrics for diffusion coefficient determination

In addition to plotting the ASSD and SSIM of the uncorrected prints against reconstructed objects at different diffusion levels, we also plotted to other common metrics: the Dice coefficient, Jaccard index, and 95$^{th}$ percentile Hausdorff distance. The former two are volumetric overlap-based indices rather than the boundary based like ASSD and return similarity scores between 0 and 1, similar to SSIM, where a higher number indicates a closer fit. The 95$^{th}$ percentile Hausdorff distance instead looks at more extreme deviations, rather than averages, returning a physical deviation like ASSD, where lower is better. These plots can be seen below in Figure S1.

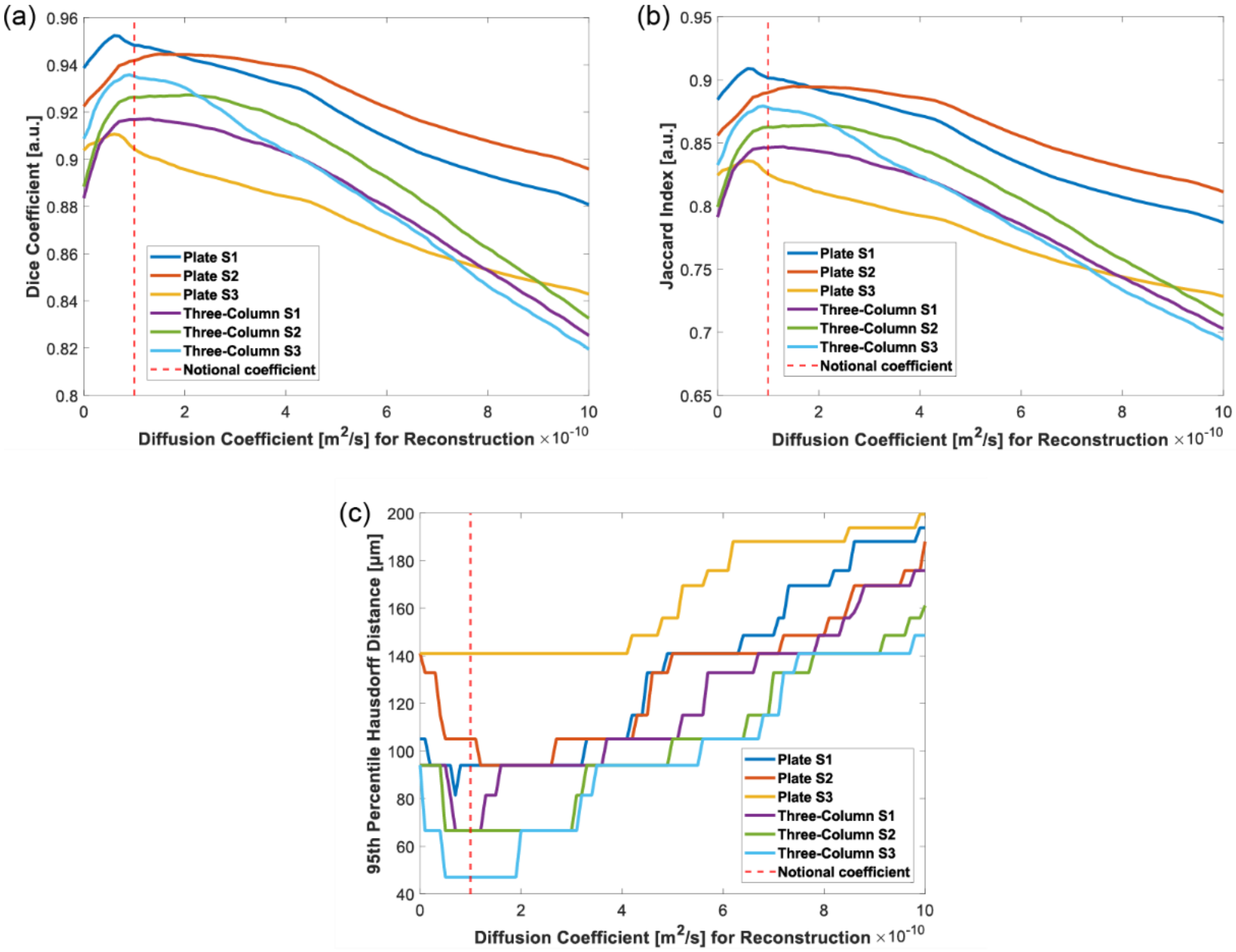

**Figure S1.** Other fidelity evaluation indices, used to determine the experimental diffusion coefficient. (a). Dice Coefficient; higher is better. (b) Jaccard Index; higher is better. (c) 95$^{th}$ percentile Hausdorff distance; lower is better.

These global metrics return similar results to the ASSD plot shown in Figure 3a. ASSD was highlighted overall due to providing deviation in physical units similar to manufacturing tolerance, while being less sensitive to noise than extremum metrics such as the Hausdorff distance.

## S.2 Print timing and kernel inputs

As our kernel takes in a time in addition to a diffusion coefficient, the approximate timing of the print must be known for optimal correction. However, different kernels may yield different print timing due to the changes in the projections.

In particular, the convolutions present in the modified Radon transform and its inverse increase the dimensions of a projection from $(N_R, N_Z)$ to $(N_R + N_{xk} - 1, N_R + N_{zk} - 1)$. This allows resulting projections to include positive intensities above and below the object, where rays would not directly cross the object. To reduce print times, CAL often (and in this work) includes a contrast maximization step due to the large number of pixels with negative intensities that are clamped to zero. For this task, we use MATLAB's 'imadjustn' function with default inputs, which maximizes intensity across the 1st and 99th percentiles of the projection set. Although the projection dimensions increase with larger kernels to allow for intensity across the full size of the object and kernel for the worst case, most of these additional pixels are zero intensity. Thus, larger kernels (and projections), lead to higher intensities in the projection and higher rates of dose delivery. 3D dose contrast is generally maintained due to the iterative optimization that follows, but since the thresholds are based on the initial projection, changes in overall intensity persist to the final optimized projections. The relative dosing rates determined by applying Eq. S7 to the reconstructions with and without diffusion at $D = 1 \times 10^{-10}\ \mathrm{m}^2/\mathrm{s}$, are depicted in Table S1.

**Table S1.** Dosing Rates from Optimized Projections

| | Plate | | Three-Column | |
|---|---|---|---|---|
| Correction Level | Without Diffusion (a.u) | With Diffusion (a.u.) | Without Diffusion (a.u) | With Diffusion (a.u.) |
| Uncorrected | 34.2 | 31.8 | 40.4 | 36.9 |
| $D = 1 \times 10^{-10}\ m^2/s$ | 38.2 | 36.9 | 51.7 | 48.7 |
| $D = 2 \times 10^{-10}\ m^2/s$ | 38.9 | 38.0 | 54.9 | 52.4 |
| $D = 3 \times 10^{-10}\ m^2/s$ | 39.4 | 42.1 | 56.8 | 54.6 |
| $D = 6 \times 10^{-10}\ m^2/s$ | 42.1 | 41.3 | 61.9 | 60.1 |
| RL Deconvolution at $D = 1 \times 10^{-10}\ m^2/s$ | 30.1 | 29.0 | 41.2 | 38.9 |

*No kernel for $D = 0$, reference used to scale input times for other correction levels

With known experimental times for the uncorrected samples, the kernel input times for the remaining levels were then determined based on the ratios of the dosing rates above. This can be an iterative process since different times will then change the resulting dosing rate, but this converges quickly as small changes in input time do not change the kernel significantly at a given diffusivity.

Since the ratios between dosing rates with and without diffusion differ, different times are obtained depending on which metric is used. Here, we accounted for both, leaning towards the dosing rates with diffusion, though any differences were within 5s. Table S2 shows the kernel input times derived from these doses, alongside the experimental print times. To properly illuminate the region from all angles, print times were rounded to the nearest 5s to cover angles in multiples of 180° under the resin's rotation rate of 36°/s. Multiples of 360° are ideal, but under minimal absorption as shown in S.2, we take 180° to be sufficient, allowing for finer control of timing.

**Table S2.** Simulated and experimental timings

| | Plate | | Three-Column | |
|---|---|---|---|---|
| Correction Level | Experimental Print Time (s) | Kernel input time (s) | Experimental Print Time (s) | Kernel input time (s) |
| Uncorrected | 130, 135 | 135* | 135, 140 | 140* |
| $D = 1 \times 10^{-10}\, m^2/s$ | 115 | 120 | 105 | 105 |
| $D = 2 \times 10^{-10}\, m^2/s$ | 115 | 115 | 100 | 100 |
| $D = 3 \times 10^{-10}\, m^2/s$ | 115 | 115 | 95 | 95 |
| $D = 6 \times 10^{-10}\, m^2/s$ | 110 | 105 | 85 | 90 |
| RL Deconvolution at $D = 1 \times 10^{-10}\, m^2/s$ | 135 | 150 | 125 | 135 |

*No kernel for $D = 0$, reference used to scale input times for other correction levels

Experimental times from shadowgram inspection differed from our kernel input times by up to 5 seconds. However, kernel variation with input time of ± 5 seconds is small in comparison to the changes in the diffusion coefficients tested. Print times in 5–10s range surrounding the ones in the table were also tested to ensure best results regarding over- and under-curing. Notably, results from the RL deconvolution method printed faster than the simulation predicted, potentially due to the higher dose delivery to the surface of the volume, which generally still form last.

While the dosing rates generally do a good job of predicting the print times at different correction levels, they do not appear to be comparable across different geometries, as the three-column structure has a higher dosing rate across all levels, both with and without diffusion, but prints in similar or even slower times than the plate for the uncorrected projections.

To decrease the effects of diffusion, it may be helpful to decrease the print times. Since the thresholds in this work were determined by applying Eq. S7 to the initial reconstructions, below we show the histogram (Fig. S2) for projections generated for a modified threshold (Eq. S1), which aimed to bring the in-part doses above that of the median in-part dose of the initial reconstruction.

$$\mu = P_{\tilde{R}*g}\left(100\% * \left(1 - \frac{\sum \frac{V_{f_{T,in-part}}}{2}}{\sum V_{f_{T,}}}\right)\right) \quad \text{(S1)}$$

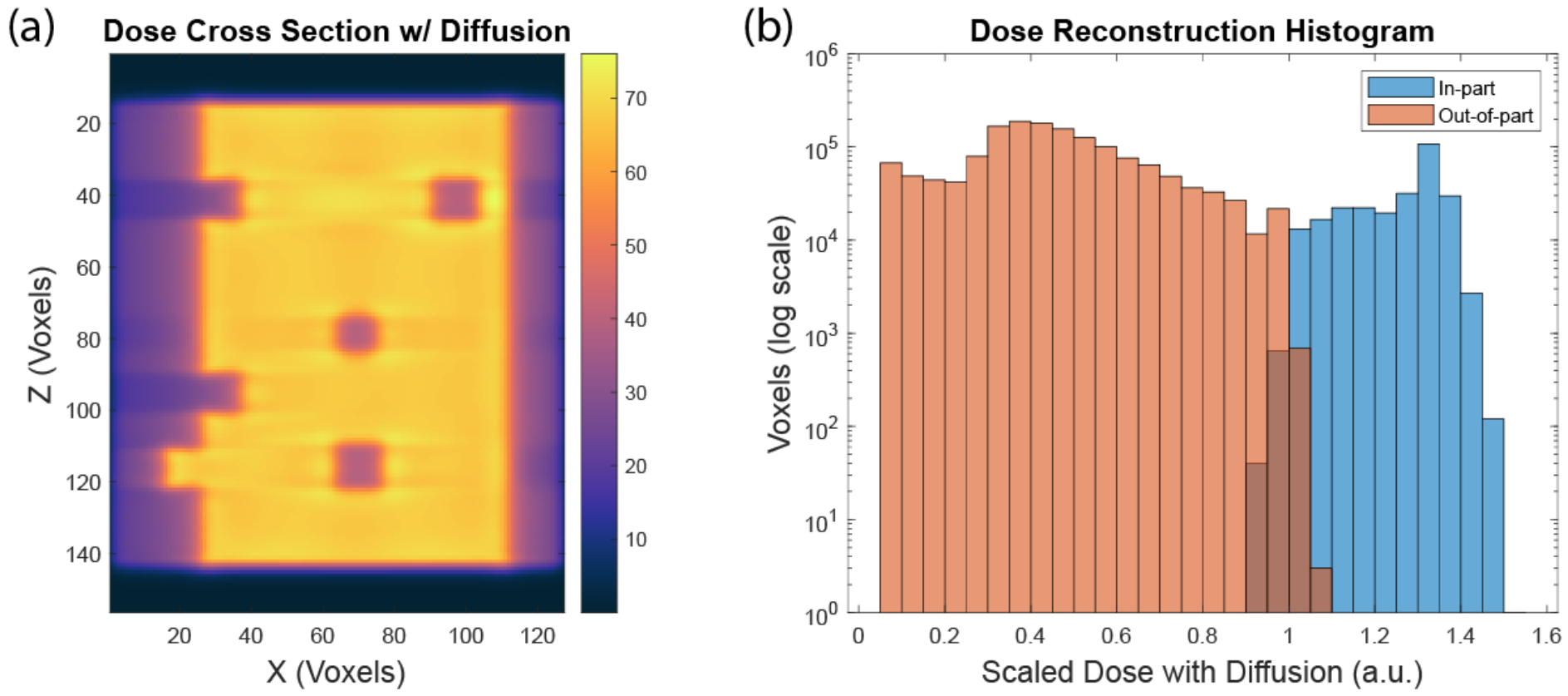

**Figure S2.** Simulated results targeting a higher dosing rate (a) Vertical cross-section of convolution-optimized dose with diffusion, optimized to target a higher threshold. (b) Histogram of voxel doses under this optimization target for $D = 1 \times 10^{-10}\ m^2/s$ .

The above figure shows the results when the optimization targeted the higher threshold dosing rate, ultimately ending with a rate of 50.63 as opposed to 36.9 (Table S1) under Eq. S7. Compared to the results in Fig. 7b and Fig 7c. the histogram overlap is comparable, with a more uniform in-part dose overall. This additionally comes with the added advantage here of printing faster due to higher dose. Beyond a certain threshold, it becomes impossible to drive doses high enough without degrading reconstruction quality. However, the exact tuning of the target threshold and relevant BCLP parameters with regard to final quality, throughput, and optimization time may be of future interest.

## S.3 Remaining experimental results

The optical micrographs of the other two samples at $D = 1 \times 10^{-10}\ \mathrm{m}^2/s$ and the one at $D = 3 \times 10^{-10}\ \mathrm{m}^2/s$ for both geometries can be seen in Figure S3.

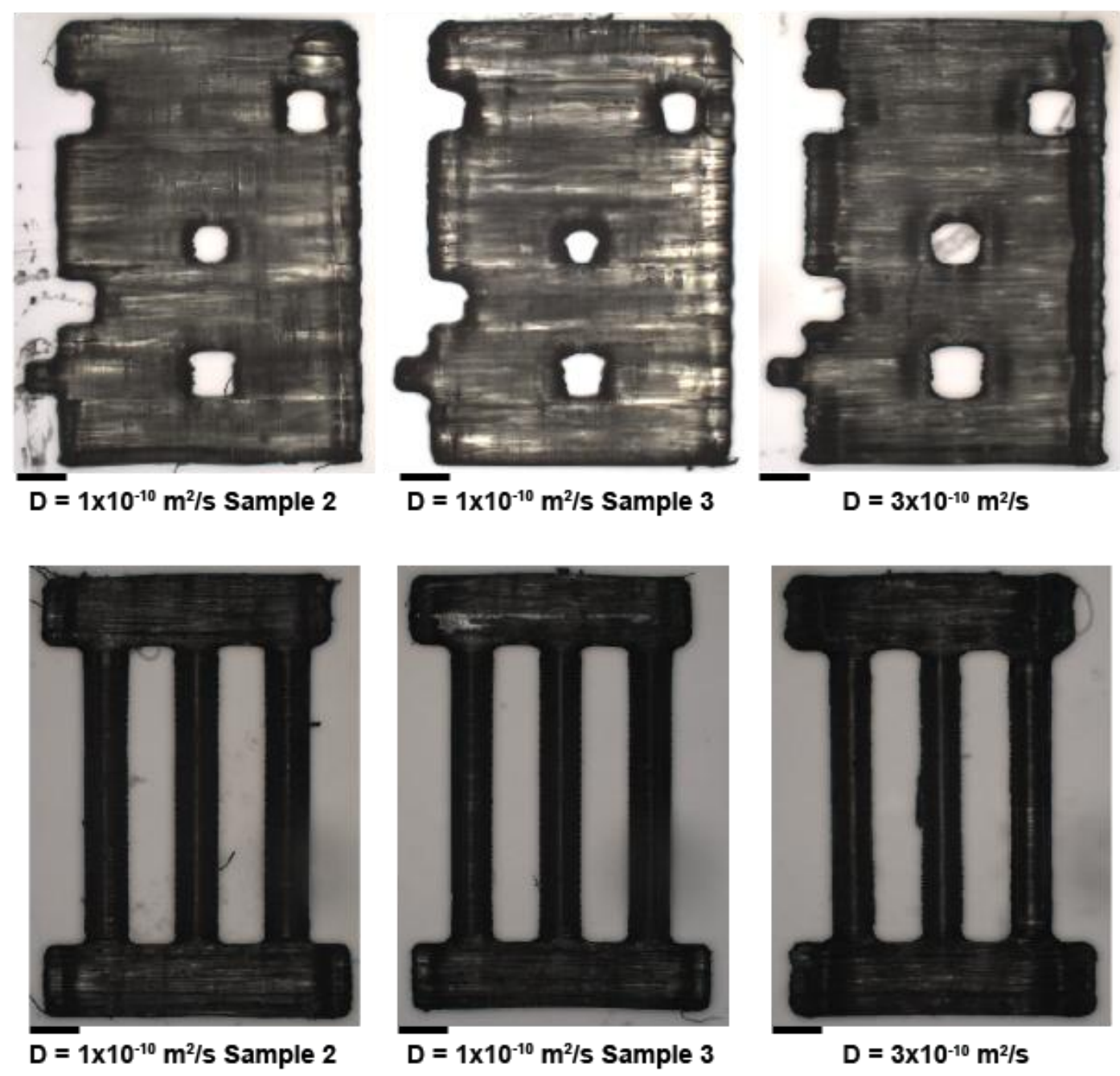

**Figure S3.** Optical micrographs of other samples. 600 µm scalebars.

Looking at these other samples, the results are rather consistent in terms of overall features with what was previously shown. The plate structure does not seem as sensitive to the correction level; all correction levels give rather good results, as mentioned in the main text. The results at $D = 1 \times 10^{-10}\ \mathrm{m}^2/s$ seem to show more square features here compared to Fig 1g, but that sample was highlighted due to higher metrics in both ASSD and SSIM, perhaps due to the specific feature size achieved or the overall bulk fidelity. As mentioned in the main text, a lack of control in the metrics for different degrees of over- or under-curing may cause some of this discrepancy between the quantitative metrics and qualitative fidelity from visual inspection. However, the consistency of the results is demonstrated here with the rest of our samples.

## S.4 Object shrinkage during photopolymerization

| **Table S3.** Shrinkage of plate geometry | |
|---|---|
| Correction Level (Sample size) | Linear shrinkage from registration (mean) |
| Uncorrected $(n = 3)$ | 17.01% |
| $D = 1 \times 10^{-10}\ m^2/s$ $(n = 3)$ | 15.44% |
| $D = 2 \times 10^{-10}\ m^2/s$ $(n = 1)$ | 15.35% |
| $D = 3 \times 10^{-10}\ m^2/s$ $(n = 1)$ | 14.52% |
| $D = 6 \times 10^{-10}\ m^2/s$ $(n = 1)$ | 13.42% |

| **Table S4.** Shrinkage of three-column geometry | |
|---|---|
| Correction Level (Sample size) | Linear shrinkage from registration (mean) |
| Uncorrected $(n = 3)$ | 13.4% |
| $D = 1 \times 10^{-10}\ m^2/s$ $(n = 1)$ | 12.0% |
| $D = 2 \times 10^{-10}\ m^2/s$ $(n = 1)$ | 10.9% |
| $D = 3 \times 10^{-10}\ m^2/s$ $(n = 1)$ | 12.7% |
| $D = 6 \times 10^{-10}\ m^2/s$ $(n = 1)$ | 10.5% |

All samples for both the plate and three-column geometries exhibited $\sim 10 - 17\%$ linear shrinkage based on the registration process of the micro-CT data to the voxelized target. Higher shrinkage was observed for the plate compared to the three-column geometry. Dimensional measurements from optical microscope images give shrinkage values $\sim 10\%$, with less variation both within and between the two geometries.

These shrinkage values are higher than commonly observed for pentaerythritol based materials ($\sim 2 - 22\%$ volume).[1,2] Previous work suggests that faster curing rates may lead to higher shrinkage during photopolymerization, a factor that may cause shrinkage to be a greater factor in CAL. Since the variations are small and shrinkage rates are rather consistent throughout for each geometry, the projections patterns can be pre-scaled to compensate for the final reduction in size. As the correction itself does not meaningfully change the shrinkage behavior, this is not a primary concern in this study, though it may warrant further investigation.

## S.5 Absorption correction

Given our parallel ray configuration, the absorption in our volume of resin can be simplified by looking at the path through a circular area. For simplicity here, we operate under the assumption that the initial intensity of light $I_0$ is equal from all angles to create a single radially defined transmission mask, similar to the approach by Madrid-Wolff et al[3]. Under the Beer-Lambert law, we have $I = I_0 e^{-\alpha l}$. For an area of radius $R$, the path length $l = rcos(\theta) + \sqrt{R^2 - r^2 \sin^2 \theta}$ where $r$ is the radius of some given point in the volume. Integrating $I$ across all $\theta$ for a constant $I_0 = 1$, we get Eq. S2.

$$I(r) = \int_0^{2\pi} e^{-\alpha(rcos(\theta)+\sqrt{R^2-r^2\sin^2\theta})}\, d\theta \tag{S2}$$

The transmission was measured using UV-VIS for different concentrations of CQ and EDAB, and the absorption coefficients were then calculated. The monomer was near-transparent (>99.9%) at 455 nm and was ignored for the absorption. EDAB had a similarly small effect at our concentrations, and CQ was measured to $3300\, m^{-1}M^{-1}$. The 2D transmission map from this effect can be seen in Figure S4.

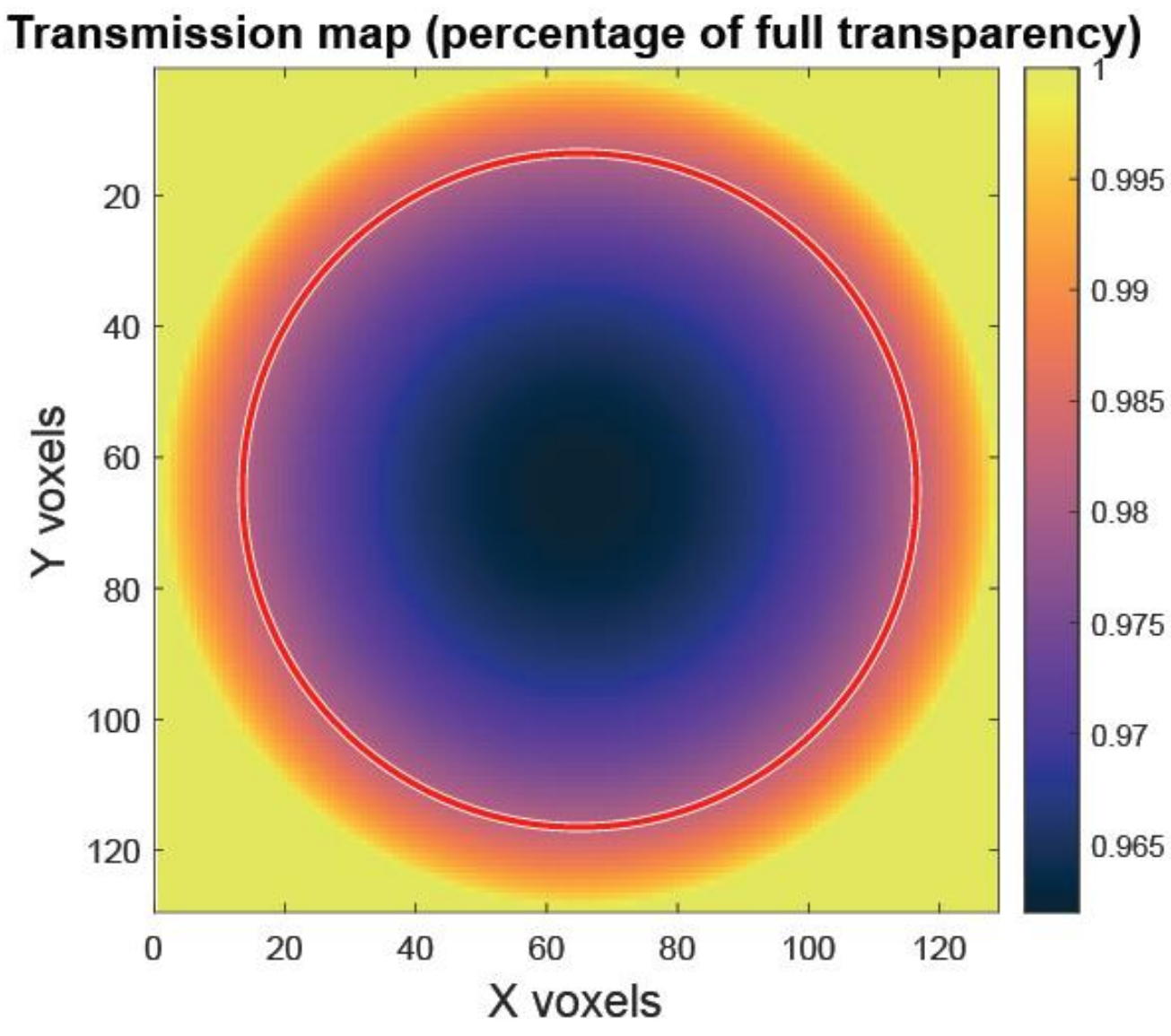

**Figure S4.** 2D transmission map for our material. 47 µm voxel size. The red circle represents the rotation path of the far edge of our largest test object, the plate geometry. The uniform intensity of 1 on the outer edges is the region outside of the print volume, our 6 mm vial.

For a 6 mm diameter vial of our material, $96.2\%$ of light reaches the center of the vial (relative to full transparency) over a full rotation. For the largest bounding box of our objects, $98.2\%$ of light reaches the point in the vial with a 2 mm radius; thus, the center of our object receives 98.0% of light from the edge.

This transmission map was added to the model and loss function gradient by modifying Eq. 5-6 to the following for use in the BCLP optimization as required.

$$\tilde{R}^* g(x, y, z) = (R^* g(x, y, z) \odot T) * K_{rq}(x, y, z) \quad \text{(S3)}$$

$$\tilde{R} f(r, \theta, z) = \int_{L(r,\theta,z)} (f(x, y, z) * K_{rq}^*(x, y, z)) \odot T ds \quad \text{(S4)}$$

where $T$ is the transmission map as shown above

$\odot$ denotes an element-wise (Hadamard) product

As mentioned above, the simplification of a constant $I_0$ from each angle creates what is essentially a radially defined transmission level. A more complete correction could be taken by simulating the dose transmission from each angle under different $I_0$ since $\int_\theta I(\theta)\, f(\theta)\, d\theta \neq I(\theta) \int_\theta f(\theta)\, d\theta$. However, this requires modification of the backprojection (adjoint Radon) implementation itself, and the above simplification is sufficient for the small amount of absorption present in our material.

Similar corrections can be made in cases where the PSF contributes significant blurring, as we could similarly modify the Radon transforms as follows.

$$\tilde{R}^* g(x, y, z) = (\int_0^\pi g(r, \theta, z) * PSF\, d\theta) * K_{rq}(x, y, z) \quad \text{(S5)}$$

$$\tilde{R} f(r, \theta, z) = \int_{L(r,\theta,z)} \left(f(x, y, z) * K_{rq}^*(x, y, z)\right) ds * PSF \quad \text{(S6)}$$

A series of 2D convolutions between the projections and a slice of the PSF along the propagation axis does assume an infinite depth of field and may be combined with the transmission correction above for more accuracy. An alternative, more complete implementation may compute all the PSF convolutions in the full 3D volume space, but as above, requires modification of the adjoint Radon implementation.

## S.6 Kernel Generation and Projection initialization

Kernels were generated through integration by Gaussian quadrature. Kernels were first generated for a full size of $2(N_x, N_y, N_z) - 1$ using one Gauss point. Rows and columns were then trimmed if the maximum value in that row or column was less than $1/\sqrt{V_{f_T}}$ $(\sim 10^{-6})$ times the maximum value of the kernel. The kernel was then recalculated for this new size using nine Gauss points for greater precision.

We generated initial guesses for the projection set using the modified Radon transform as described in Eq. 6, followed by the Ram-Lak filter and non-negative clamping. Since this involves taking the Radon transform on the target after convolution by the diffusion kernel, it expectedly causes the initial guess to create an even more blurred reconstruction than the standard FBP. While this was chosen for simplicity of consistency with the rest of the model, the gradient descent method used here can be sensitive to the initial guess, which can be chosen in a variety of ways. Table S5 compares the output results for initial guesses derived from the modified Radon transform, the standard Radon transform, and a Radon transform conducted on a target after RL deconvolution with the kernel.

**Table S5.** Results for the three-column structure for different initializations

| Initialization method | Kernel input time (s) | Iterations to convergence | Dosing rate without diffusion (a.u.) | Dosing rate with diffusion (a.u.) | Simulated ASSD (µm) | Simulated SSIM (a.u.) |
|---|---|---|---|---|---|---|
| Modified Radon | 105 | 77 | 51.73 | 48.74 | 1.17 | 0.9970 |
| Standard Radon | 110 | 131 | 50.74 | 45.73 | 1.01 | 0.9973 |
| Deconvolved Target | 135 | 193 | 40.68 | 39.97 | 1.54 | 0.9965 |

The results here are similar for the three initialization methods, with differences largely due to the dosing rate changes between the methods yielding different input kernels (S.2.) However, with kernels equalized, there are still notable—though minor—differences in output for these simple geometries. To further improve the projection generation process, investigation into initialization methods, or ways to get the optimization to explore more global solutions may be of interest in the future.

## S.7 Band Constraint $L_p$ norm (BCLP) optimization implementation and parameters

The loss function and gradient defined in Eq. 7-8 require a target dose, ideally without normalization as to maintain scaling between iterations[4]. In the past, this has been set as a proportion (often 80% to 95%) of the maximum intensity of the initial reconstruction[5,6].

Here, we instead define the target dose based on a percentile value of the initial reconstruction, with the idea to reduce noise and target a threshold closer to the initial guess. Specifically, the threshold is determined by applying Eq. S7 to the initial reconstruction.

$$\mu = P_{\tilde{R}*g}(100\% * (1 - \frac{\sum V_{f_{T,in-part}}}{\sum V_{f_{T,}}})) \tag{S7}$$

Where $P_{\tilde{R}*g}$ is the percentile function of the initial dose or repletion reconstruction

$V_{f_T}$ is the voxel count of the target matrix, equal to $N_x N_y N_z$

$V_{f_{T,in-part}}$ is the number of in-part voxels

If the reconstruction perfectly separated in- and out-of-part regions, this threshold would find the lowest dose of any in-part voxel. More generally, this represents the threshold that cures a region of equal volume to our target. The high and low target bands are then scaled accordingly to this threshold $\mu$ (S2, S3).

$$t_h = (1 + \delta + \varepsilon \pm \varepsilon)\mu$$
$$t_l = (1 - \delta - \varepsilon \pm \varepsilon)\mu$$

where $t_h$ and $t_l$ are the high and low target bands ($[1.2\mu, 2.0\mu]$ and $[0, 0.8\mu]$ in this work)

$\delta$ is the half-width of the desired separation between the bands ($0.2$ in this work)

$\varepsilon$ is the half-band width of the tolerance for the in and out of part regions ($0.4$ in this work)

The standard 2-norm raised to the first power was used in the optimization as seen in Eq 6-7. Weights were set to a constant $1/\sqrt{V_{f_T}}$ for all voxels, with scaling by the inverse square root of the number of elements in the volume to maintain the scale of the 2-norm. Exit conditions were set for when the change in the loss function differed by less than some tolerance times the previous loss function (Eq. S8).

$$\frac{\|\mathcal{L}(x_n) - \mathcal{L}(x_{n-1})\|}{\mathcal{L}(x_{n-1})} < \tau \tag{S8}$$

Adaptive step sizes were applied using spectral gradient descent via the geometric mean of the two Barzilai-Borwein (BB) step sizes[7]. This provides a simple method of updating the step size using only the norm of the loss function at the current and previous steps, requiring persistent storage of only a single scalar between iterations. For stability the step sizes were clamped to be non-increasing for any steps beyond the second (Eq. S9), with the second step allowed to increase in case the user specified initial step size is too small. We used an initial step size of 1, noting that the loss function and gradient magnitude related to the weights (uniformly $1/\sqrt{V_{f_T}}$ here).

$$\alpha_1 = \frac{\alpha_0 \|\nabla\mathcal{L}(x_n)\|}{\|\nabla\mathcal{L}(x_{n+1}) - \nabla\mathcal{L}(x_n)\|}$$

$$\alpha_{n+1} = \alpha_n * \min\left(1, \frac{\|\nabla\mathcal{L}(x_n)\|}{\|\nabla\mathcal{L}(x_{n+1}) - \nabla\mathcal{L}(x_n)\|}\right), n > 0 \qquad \text{(S9)}$$

An example plot of both the loss function and the step size is shown in Fig S5.

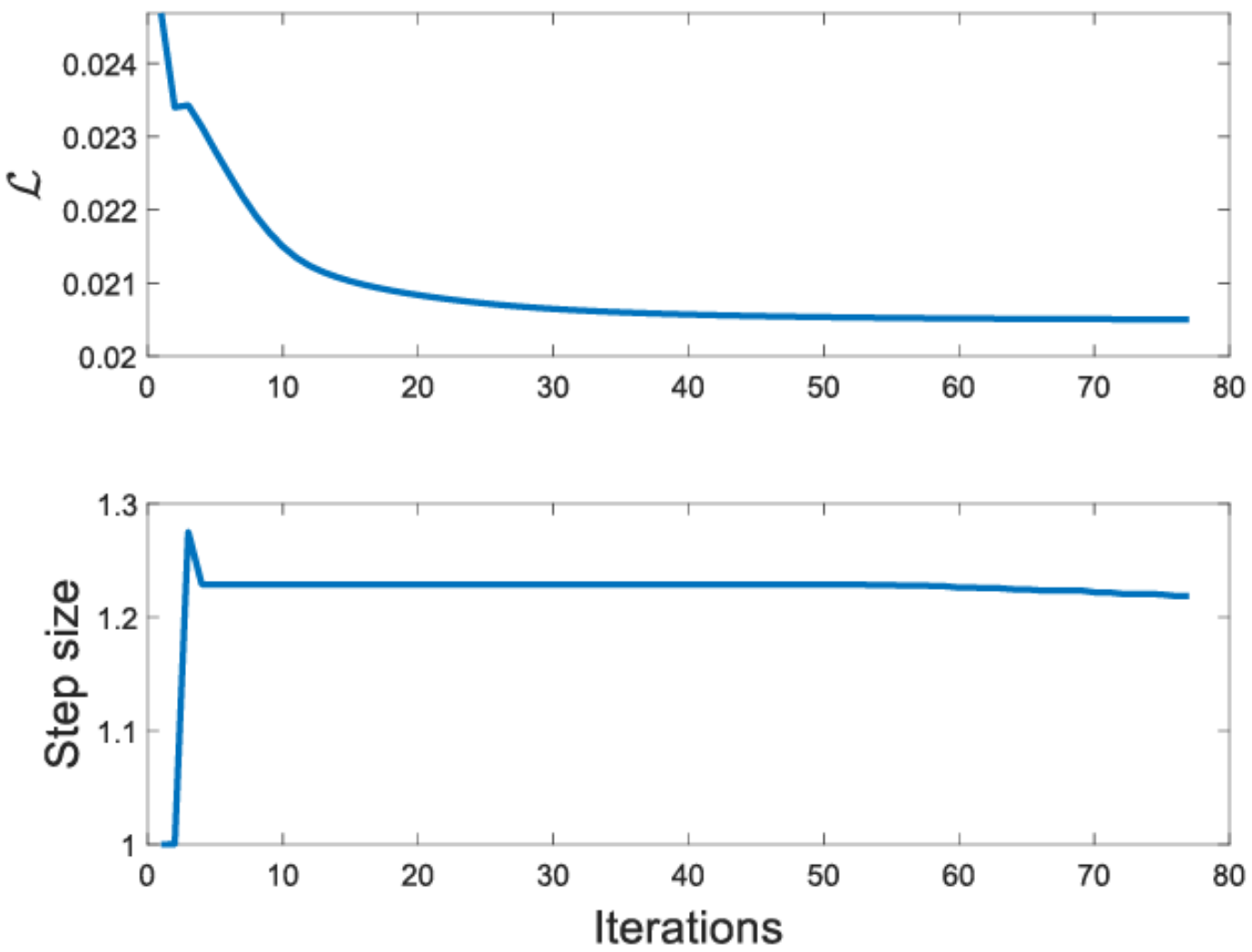

**Figure S5.** Loss function and step size for multiple iterations of the three-column geometry corrected for $D = 1 \times 10^{-10}\ m^2/s$.

## Supplemental References